\documentclass[letterpaper, 10 pt, conference]{ieeeconf} 

\IEEEoverridecommandlockouts                     
\title{\LARGE \bf Mining Threat Intelligence about Open-Source Projects and Libraries from Code Repository Issues and Bug Reports
}

\author{Lorenzo Neil, Sudip Mittal, and Anupam Joshi \\
University of Maryland, Baltimore County, Baltimore, MD 21250, USA\\
Email: $\lbrace$lneil1, smittal1, joshi$\rbrace$@umbc.edu
}

\usepackage{url} 
\usepackage{graphicx}
\usepackage{float}
\usepackage{algorithm}
\usepackage{algorithmic}
\usepackage{tablefootnote}
\usepackage{listings}
\usepackage{color}
\usepackage{listings,multicol}  
\usepackage{filecontents}    

\definecolor{dkgreen}{rgb}{0,0.6,0}
\definecolor{gray}{rgb}{0.5,0.5,0.5}
\definecolor{mauve}{rgb}{0.58,0,0.82}
\begin{document}

\maketitle
\thispagestyle{empty}
\pagestyle{empty}

\begin{abstract}
Open-Source Projects and Libraries are being used in software development while also bearing multiple security vulnerabilities. This use of third party ecosystem creates a new kind of attack surface for a product in development. An intelligent attacker can attack a product by exploiting one of the vulnerabilities present in linked projects and libraries.

In this paper, we mine threat intelligence about open source projects and libraries from bugs and issues reported on public code repositories. We also track library and project dependencies for installed software on a client machine. We represent and store this threat intelligence, along with the software dependencies in a security knowledge graph. Security analysts and developers can then query and receive alerts from the knowledge graph if any threat intelligence is found about linked libraries and projects, utilized in their products.

\end{abstract}


\section{INTRODUCTION}
In the normal course of software development, developers and coders rely on various open source projects and libraries. Open source projects and libraries comprise of source code that is open for anyone to inspect, modify, update or enhance \cite{opensourcecom}. This type of programming embraces innovative ideas of collaboration, open exchange, transparency, and community-oriented development. Software developers and programmers are then able to access this source code easily and add their input of code or fix parts that may not function as intended. This is much different from ``proprietary'' or ``closed source'' software which only allows access and modifications from the person, group, or organization that own it. Also, closed source software requires users to sign a license and accept the terms placed under it by the originators which differs drastically from legal terms of open source software \cite{vaughannichols_2015}. The  design of open source projects and their licenses encourages all computer programmers to access public projects and collectively make edits however they find fit.

This use of third party ecosystem creates a new kind of attack surface for the product in development. The developers link these products and code libraries with little consideration of threats, vulnerabilities, and exposures present in them. It is estimated that about 70-80\% of source code implemented in current day projects are from open source communities \cite{vaughannichols_2015}. The abundance of open source projects also increases the need to track security vulnerabilities and bugs that are present within the libraries they are linked to. Such security vulnerabilities are then inherited by the product in development. An intelligent attacker can then attack the product by exploiting one of these vulnerabilities. 
The problem gets more complex with the number of recorded open source vulnerabilities each year \cite{opensourcecom}. Another problem that even compounds the issue is the fact that the open source libraries linked by a developer can themselves link to other vulnerable libraries and so on. A vulnerability present in any of these can cause a domino effect, resulting in the product being developed inheriting a vulnerability.

A recent case would be the exploitation by hackers on an open source security vulnerability within Equifax's, a U.S. based credit rating bureau, database during July - August 2017 \cite{opensourcecom}. In this attack, it was reported that hackers attacked Equifax by exploiting a web applications vulnerability to access private files of customers. The reported breach stemmed from the popular open source programming framework that Equifax utilized for their web applications known as Apache Struts \cite{opensourcecom}. The Apache Struts framework included a critical vulnerability that allowed easy access for hackers to Equifax's database. Names, social security numbers, birth dates, addresses, and driver’s license numbers from an estimated 143 million customers were compromised \cite{opensourcecom}. Equifax is not alone in the sense that other companies also have implemented known vulnerable open source components into their organization and products. This implementation of open source software with persistent vulnerabilities is a growing concern for software developers. 

In order to better protect the product in development, it is necessary to create a repository of known vulnerabilities in these open source libraries and projects. Threat intelligence about some of these projects can be mined using {\em traditional} sources like NIST's National Vulnerability Database (NVD)\footnote{\url{https://nvd.nist.gov/}}, United States Computer Emergency Readiness Team (US-CERT)\footnote{\url{https://www.us-cert.gov/}}, etc. Other sources which are more {\em non-traditional} are, Twitter, Reddit, blogs, and news. Non-traditional sources are faster than the traditional ones. There is a significant gap between initial vulnerability announcement and NVD release \cite{attack2017register}. Vulnerability threat intelligence appears first on non-traditional sources \cite{vul2017register}. Mining non-traditional sources is becoming really important. In our previous work, we have developed \emph{CyberTwitter} \cite{mittal2016cybertwitter} and \emph{Cyber-All-Intel} \cite{mittal2017thinking} systems that mines threat intelligence from various OSINT sources. The systems then represent cybersecurity intelligence in knowledge graphs and vector spaces so it can be used by artificial intelligence based cyber-defense systems.

\begin{figure}[h]
\centering
\includegraphics[scale=0.4]{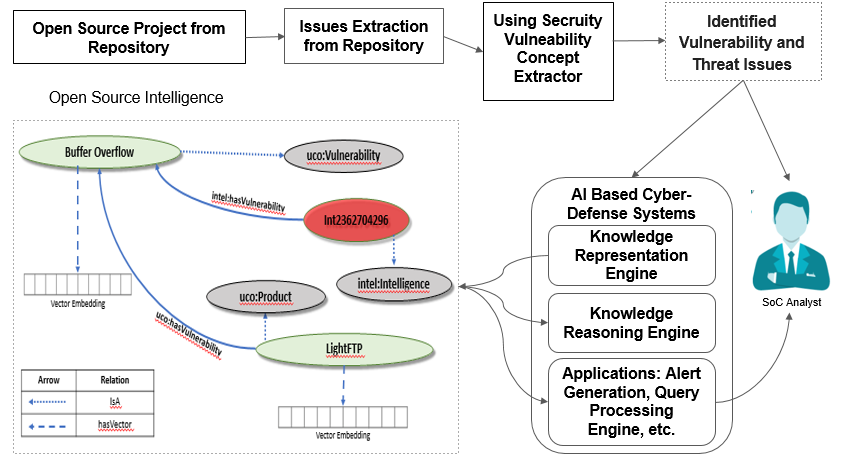}
\caption{Mining threat intelligence from bug and issue reports. Intelligence is stored in a Security Knowledge Graph. The graph represents the threat intelligence: ``\emph{I've noticed a buffer overflow in the Unix version of LightFTP v1.1}''}
\label{fig:overview}
\end{figure}

In this paper, we propose a `shift-left' \cite{ryanlittlefield_2018} form of security. We create a system that will inform a developer about potential threats and vulnerabilities that the product in development might inherit as a result of linking to a vulnerable open source project or library. We mine threat intelligence from issues and bugs raised on web-based hosting service for version control like, GitHub \cite{github}, GitLab \cite{gitlab}, bitbucket \cite{bitbucket}, etc. These platforms have been used by developers to host and collaborate on source code development \cite{thenewstack_2018}. We extract vulnerabilities raised on these platforms and represent them in a security knowledge graph. The knowledge graph then becomes a store for various vulnerabilities and exposures present in various open source projects, products and libraries (see Figure \ref{fig:overview}). This knowledge can then be queried by various developers helping them create products that are secure from the ground up. Pervasive software security entails cyber supply chain risk management, where the developers are made aware of various threats present in libraries they link with the development of their products.

We also create another application that can track installed software on a client machine, and then use the above mentioned knowledge graph to \emph{reason} alerts for the security analyst. These alerts warn the analyst if an installed software is linked to a vulnerable open source library or project. We built a proof of concept for this application that runs on a linux installation.



The rest of the paper is organized as follows - Section \ref{relwork} discusses some of our related work. Our methodology is presented in Section \ref{method}. We discuss our experimental setup and evaluation in Section \ref{results}. We conclude in Section \ref{conc}.

\section{Related Work}\label{relwork}

\subsection{Open Source Software Security}
Open source development 
has created a variety of new software security challenges. Closed source proponents claim that the availability of open source software's code allows hackers to easily find a way to compromise the security. They believe that ``hackers finding the source code and placing back doors for unauthorized access to their systems is one of the biggest limiting factors for open source software'' \cite{989921}. 
Closed source systems based on the principle of security through obscurity, may have theoretical or actual security vulnerabilities, but its owners or designers believe that they are more secure if the flaws are not known.

Security of open source systems stems from the Kerckhoffs's Law (sometimes refereed to as Shannon's maxim), which states that "A cryptosystem should be designed to be secure if everything is known about it except the key information." \cite{shannon1949communication}. The strength behind open source software stems from it's wide audience looking at the code and collectively finding problems. In our work, we are interested in gathering threat intelligence about issue and vulnerability reports in open source repositories from project contributors. 

\subsection{Mining Threat Intelligence}
The production of new computer systems and software has always attracted the likes of cyber criminals who have compromised such technology through hacking or infecting with viruses and/or malware. 
Their objectives can range from shutdown of a website, data breaches, acts of fraud, or distribution of viruses. However, understanding the volume of these cyber attacks and all the broad spectrum of vectors in which they occur greatly aids cyber security professionals effort to create cyber-defense strategies. 
The exponential growth of hacking communities among the web also means that security professionals also gain an exponential amount of threat intelligence from those communities and are tasked with a challenge to monitor their behavior. The threat intelligence that comes from these communities can be forums consisting of numerous members, posts, or threads that require constant filtering and mining for security vulnerabilities \cite{ryanlittlefield_2018}. 

The use of semantic knowledge graphs in cybersecurity has gained traction in the past few years. Considerable attention has been dedicated to develop techniques for extracting concepts related to security vulnerabilities, affected software, hardware, and organizations and generating its semantic representation~\cite{mulwad2011extracting}\cite{JoshiNER2013}
\cite{syed2015uco}
\cite{mittal2016cybertwitter}\cite{mittal2017thinking}. While previous research focused on sources such as NVD, social media, and security blogs, our work is applied to bug and issue reports taken from public code repositories, where the content is different from other sources.


\subsection{Knowledge Graph systems for Threat Intelligence}

Effective strategies to precisely counter cyber attacks relies heavily on the detection of future threats, attacker's behavior, and accessibility of threat intelligence. Being able to attain cyber threat intelligence will eliminate the possibility of vulnerabilities being exploited and improves the ability to react to future attacks \cite{iannacone_2015}. 

Knowledge graphs for cybersecurity have been used by \cite{kandefer2007symbolic} to create association and combination of data and information from multiple sources. 
Takahashi et al. \cite{takahashi2010ontological,takahashi2010building} built an ontology for cybersecurity operational information based on actual cybersecurity operations mainly focused on cloud computing-based services. Rutkowski et al. \cite{rutkowski2010cybex} created a cybersecurity information exchange framework, known as CYBEX. Another insightful work by Xie et al. \cite{xie2010using} discusses uncertainty modeling for cyber security centered around near real-time security analysis such as intrusion response. 

Intrusion detection and prevention systems (IDPS) help by examining signature markings of cyber infrastructure for malicious activity and generating alerts. These systems however cannot detect malicious activity if the systems do not already have that signature in their databases \cite{mittal2016cybertwitter, more_matthews_joshi_finin_2012}. Also, these systems are point source solutions and cannot organize threat intelligence coming from heterogeneous sources. 
Cyber research has focused on combining traditional signature based intrusion detection with ontological reasoning to further improve knowledge graphs. This allows for the interpretation of means and consequences for links between cyber threats and vulnerabilities whose signatures are not yet in the databases. 

\section{Methodology}\label{method}

\begin{figure}[h]
\centering
\includegraphics[scale=0.55]{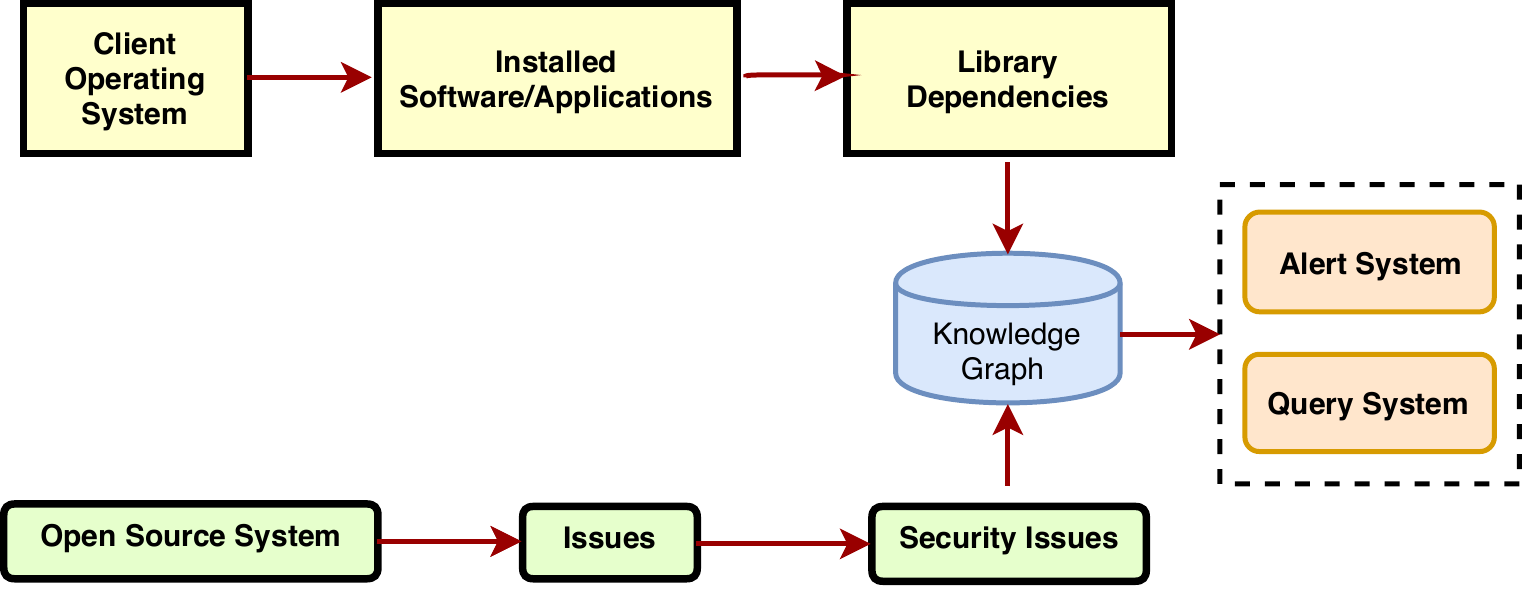}
\caption{Knowledge Graph with ontology of projects and threat intelligence.}
\label{fig:arch}
\end{figure}

In this section, we will describe our system architecture (see Figure \ref{fig:arch}). Our system has a two subparts. The first part tracks repositories of various open source projects and libraries. It monitors various issues and bugs reported on these repositories and filters out all security related issues that mention vulnerabilities, threats, etc. It then converts this intelligence into a machine readable format. The second part tracks software running on a machine and identifies all libraries and dependencies. Both parts feed the collected information to a common security knowledge graph. 

Once the security knowledge graph gets populated, a system administrator or a developer can use a query engine and an alert generation system to find vulnerable projects and libraries. We now describe each of the different subparts and applications in detail: 

\subsection{Initializing a Security Knowledge Graph}\label{kg}
Our knowledge graph consists of vulnerabilities, threats found on open source projects, library repositories, and installed software with their dependencies. For our knowledge graph we used the Unified Cybersecurity Ontology (UCO) \cite{syed2015uco} to provide cybersecurity domain knowledge. An Intelligence ontology \cite{mittal2016cybertwitter} was used to represent threat intelligence. We also created a \emph{software dependency} ontology that helps represent installed software and its dependencies (see Section \ref{deponto}). In our system, we also  matched entities of various softwares and libraries that we encounter to their DBpedia \cite{auer2007dbpedia} equivalent entities. For example the installed software \emph{vlc} is matched to \emph{dbr:VLC\_media\_player}, this helps us retrieve more global information like, developer, genre, operating system, etc. about a particular software. 

\subsection{Tracking Installed Software \& Mining Library Linking}\label{deponto}

To track and generate alerts for all software that inherits vulnerabilities from open source dependencies, we compiled the knowledge of all installed software on a machine and its dependencies. We obtained all of the software programs already installed on a machine and then for each program we listed all of the library dependencies and environment variables. The environment variables helped us identify open source projects that are utilized by a particular installed software. In order to create a proof of concept, we focused on a linux installation. We utilized the \emph{objdump}\footnote{\url{https://linux.die.net/man/1/objdump}} tool to disassemble installed software and trace dependencies. Once we obtained the information we asserted it in our security knowledge graph mentioned in Section \ref{kg}, using the software dependency ontology. 

\begin{figure}[h]
\centering
\includegraphics[scale=0.5]{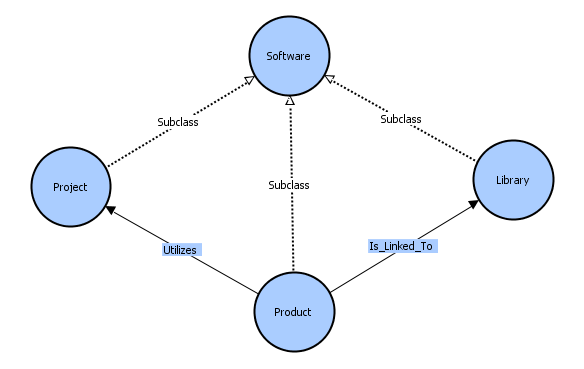}
\caption{Software Dependency Ontology Schema.}
\label{fig:figure1}
\end{figure}

An abstract view of the software dependency ontology is show in Figure \ref{fig:figure1}. The software class is divided into three sub classes which are the project, product, and library. The product class represents installed software on a machine. A product \emph{utilizes} a project, and \emph{is linked to} a library.  


\begin{figure}[!htbp]
\small
\begin{minipage}{1\columnwidth} 

@prefix uco: $<$http://accl.umbc.edu/ns/ontology/uco\#$>$ . \\
@prefix soft: $<$http://accl.umbc.edu/ns/ontology/software\#$>$ . \\
@prefix rdf: $<$http://www.w3.org/1999/02/22-rdf-syntax-ns\#$>$ . \\
@prefix rdfs: $<$http://www.w3.org/2000/01/rdf-schema\#$>$ . \\
@prefix xml: $<$http://www.w3.org/XML/1998/namespace$>$ . \\
@prefix xsd: $<$http://www.w3.org/2001/XMLSchema\#$>$ . \\
@prefix dbp: $<$http://dbpedia.org/resource\#$>$ . \\
@prefix owl: $<$http://www.w3.org/2002/07/owl\#$>$ . \\
\\
$<$/usr/bin/python3.6$>$ a soft:Product ; \\
    soft:Is\_Linked\_To $<$libutil.so.1$>$ ;\\
    soft:Is\_Linked\_To $<$libpython3.6m.so.1.0$>$ ;\\
    soft:Is\_Linked\_To $<$libm.so.6$>$ ;\\
    soft:Is\_Linked\_To $<$libdl.so.2$>$ ;\\
    soft:Is\_Linked\_To $<$libpthread.so.0$>$ ;\\
    soft:Is\_Linked\_To $<$libc.so.6$>$ .\\
\\
$<$libutil.so.1$>$ a soft:Library .\\
$<$libpython3.6m.so.1.0$>$ a soft:Library .\\
$<$libm.so.6$>$ a soft:Library .\\
$<$libdl.so.2$>$ a soft:Library .\\
$<$libpthread.so.0$>$ a soft:Library .\\
$<$libc.so.6$>$ a soft:Library .\\

\end{minipage}
\caption[RDF example.]{RDF for libraries linked by the program Python 3.6.}
\label{fig:rdfpython}
\end{figure}

\subsection{Bug \& Issue Tracking}\label{isstrack}
\begin{figure}[h]
\centering
\includegraphics[scale=0.6]{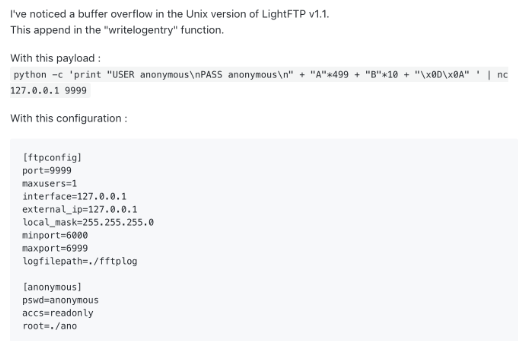}
\caption{Sample issue showing security buffer overflow in a popular Unix FTP client.}
\label{fig:samplevul}
\end{figure}

So as to mine active threat intelligence for various open source projects and libraries, we collected bug reports and issues posted by developers on code repositories. We then asserted the information in our security knowledge graph (see Section \ref{assert}). Figure \ref{fig:samplevul}, shows a vulnerability in a popular FTP client for Unix. For our system, as a proof of concept, we have developed a crawler and tracker to collect threat intelligence from repositories hosted on GitHub \cite{github}. We utilized GitHub's Rest API to collect and track open issues, closed issues, and pull requests from a project repository. 

We retrieved all the issues present in multiple project repositories in JSON format and stored them in data frames. The resulting data frames for each project were then filtered to remove any non-security related issues. We extracted issues that consisted of terms related to security vulnerabilities using a \emph{Security Vulnerability Concept Extractor} (SVCE) \cite{mittal2016cybertwitter}. The SVCE is able to tag every sentence with the following concepts: Means of an attack, Consequence of an attack, affected software, hardware and operating system, version numbers, network related terms, file names and other technical terms.

SVCE discards all issues for which it fails to identify at-least two security concept, thus further removing non-security related data. The extracted concepts are also used to generate an RDF Linked Data representation for every issue that may be queried by security systems to protect against potential attacks. 


\subsection{Asserting Threat Intelligence}\label{assert}

Once the SVCE identifies security concepts and entities. We then associated them with Uniform Resource Identifiers (URIs). These URIs are then converted to nodes in our security knowledge graph. We used the Unified Cybersecurity Ontology \cite{syed2015uco} which integrates heterogeneous data and knowledge schemas from different cybersecurity systems and standards.

We used DBpedia to link various knowledge graph nodes to real world concepts. Entity matching
process is performed by using DBpedia \cite{auer2007dbpedia} and DBpedia spotlight \cite{mendes2011dbpedia}. For example we can use DBpedia to map the string ``Adobe Flash'' to \emph{dbr:Adobe\_Flash}. This external knowledge graph help us map our entities to real world conceptual instances. 

After entity linking, we stored the linked data as RDF triples in our security knowledge graph. In our system we need information of cybersecurity intelligence. Threat intelligence is temporal in nature and may contain other meta-data like, origin, credibility, provenance, etc. UCO though gives us a domain overview of cybersecurity it cannot handle temporal nature of events. So as to handle time in events we
use the intelligence ontology \cite{mittal2016cybertwitter}. 

To give an example, Figure \ref{fig:exampleRDF} shows the RDF statements created for the intelligence ``\emph{I've noticed a buffer overflow in the Unix version of LightFTP v1.1}'' (Figure \ref{fig:samplevul}). A graphical representation of the same intelligence, ‘Int2362704296’ has been shown in Figure \ref{fig:overview}. 
Once we obtain the intelligence in the RDF format, we use it to create the alert and the query system.

\begin{figure}[!htbp]
\small
\begin{minipage}{1\columnwidth} 

@prefix uco: $<$http://accl.umbc.edu/ns/ontology/uco\#$>$ . \\
@prefix intel: $<$http://accl.umbc.edu/ns/ontology/intelligence\#$>$ . \\
@prefix rdf: $<$http://www.w3.org/1999/02/22-rdf-syntax-ns\#$>$ . \\
@prefix rdfs: $<$http://www.w3.org/2000/01/rdf-schema\#$>$ . \\
@prefix xml: $<$http://www.w3.org/XML/1998/namespace$>$ . \\
@prefix xsd: $<$http://www.w3.org/2001/XMLSchema\#$>$ . \\
@prefix dbp: $<$http://dbpedia.org/resource\#$>$ . \\
@prefix owl: $<$http://www.w3.org/2002/07/owl\#$>$ . \\
\\
$<$Int2362704296$>$ a intel:Intelligence ; \\
    intel:hasVulnerability $<$buffer\_overflow$>$ .\\
\\
$<$LightFTP$>$ a uco:Product ;\\
	uco:hasVulnerability $<$buffer\_overflow$>$ ;\\
    owl:sameAs dbp:FTP-server .\\
\\
$<$buffer\_overflow$>$ a uco:Vulnerability ;\\
    uco:affectsProduct $<$LightFTP$>$ ;\\
    owl:sameAs dbp:buffer\_overflow .\\
\end{minipage}
\caption[RDF example.]{RDF for textual input ``I've noticed a buffer overflow in the Unix version of LightFTP v1.1''. Also, $owl:sameAs$ property has been used to augment the data using an external source `DBpedia' \cite{auer2007dbpedia}.}
\label{fig:exampleRDF}
\end{figure}

\subsection{Applications}

Once we populated our security knowledge graph with the information about installed software and linked dependencies (see Section \ref{deponto}) and also add, threat intelligence mined from issues and bug reports (Section \ref{assert}), we utilized it to create an alert generation system and a query system.

\subsubsection{Query System}
The developer before linking to an open source library or using a project should be able to query the security knowledge graph to check for known vulnerabilities. We have created a SPARQL\footnote{\url{https://www.w3.org/TR/rdf-sparql-query/}} endpoint that can accept queries which run on our knowledge graph. An example query to list all vulnerabilities in \emph{LightFTP}: 

\begin{lstlisting}
SELECT ?y WHERE {
?LightFTP <hasVulnerability> ?y .
}
\end{lstlisting}

An example query to look up vulnerabilities in linked libraries to the installed application \emph{firefox}: 

\begin{lstlisting}
SELECT ?x WHERE {
?firefox <Is_Linked_to> ?z
?z <hasVulnerability> ?x.
}
 \end{lstlisting}

\subsubsection{Alert Generation System} The system will reason on our security knowledge graph and generate alerts. In our system we include SWRL rules\footnote{\url{https://www.w3.org/Submission/SWRL/}} so as to generate alerts. SWRL rules contain two parts, antecedent part (body), and a consequent (head). The body and head consist of conjunctions of a set of ‘atoms’. Informally, a rule may be read as meaning that if the antecedent holds (is “true”), then the consequent must also hold. For our system we see two potential alert scenarios: 
\begin{enumerate}
\item A developer is linking to a library or a project with known vulnerabilities and threats: The system will take in all the libraries that a developer wants to use and then trigger an alert if it finds a vulnerability or threat in any one of these libraries. We see this as a developer initiated scenario. 

Our alert system also checks other linked libraries that link to the ones mentioned by the developer. Some example rules included in our system: 

\begin{lstlisting}
Rule for vulnerable project utilization: 

Product(?x)^ Utilizes(?x, ?y)^ hasVulnerability(?y, ?z) 
==> RaiseAlert(?x,"Yes")

Rule for linked library vulnerability check:

Product(?x)^ Is_Linked_To(?x, ?y)^ hasVulnerability(?y, ?z) 
==> RaiseAlert(?x,"Yes")

Rule for secondary linked library vulnerability check:

Product(?x)^ Is_Linked_To(?x, ?y)^ Is_Linked_To(?y, ?z)^ hasVulnerability(?z, ?u) 
==> RaiseAlert(?x,"Yes")

\end{lstlisting}

The rules raise an alert if any vulnerability or threat is found in linked libraries and projects. For the first rule above `Rule for vulnerable project utilization', given a product node $?x$, the rules check for edge: $Utilizes(?x, ?y)$, to hop to the graph node $?y$. Once at $?y$ it checks for the edge: $hasVulnerability(?y, ?z)$ to hop to node $?z$. If the above node exists an alert is generated. A similar technique is used to evaluate other rules mentioned above. 

\item An installed application on a client machine is linked to compromised dependencies: This is an information triggered alert, where influx of new threat intelligence warrants a lookup for vulnerable installed software. The system should automatically inform a security analyst that an installed application on a client machine is vulnerable. An example rule for this alert: 

\begin{lstlisting}
Rule for vulnerable libraries: 

Library(?x)^ hasVulnerability(?x, ?y)^ Is_Linked_To(?z, ?x)
==> RaiseAlert(?z,"Yes")

Rule for vulnerable projects: 

Project(?x)^ hasVulnerability(?x, ?y)^ Utilizes(?z, ?x)
==> RaiseAlert(?z,"Yes")

\end{lstlisting}

\end{enumerate}

\section{Experimental Setup \& Evaluation}\label{results}

In order to evaluate the system and collect empirical data we ran our system under experimental conditions. The system was run on a Ubuntu\footnote{\url{https://www.ubuntu.com/}} Linux installation with 81 installed programs, some of these were pre-installed. We extracted the library and project dependencies for these 81 installed programs and represent this information in our security knowledge graph (see Section \ref{deponto}). Figure \ref{fig:rdfpython}, shows the triples generated for a popular installed software. 

We collected 110,800 issues posted on GitHub \cite{github}, using the GitHub Rest API. For our experiments and to create a valid proof of concept, we limit issue collection to the GitHub repositories for the 81 installed projects. We also use only the issues posted after January 2018 in our analysis. Out of the 110,800 issues collected our SVCE (see Section \ref{isstrack}) filtered 9,194 security issues. We then assert these security vulnerabilities in our knowledge graph (see Section \ref{assert}). Figure \ref{fig:exampleRDF}, lists triples generated for a popular FTP client. 

We performed an initial evaluation of our prototype system using the bug-reports collected.  We evaluate the quality of the tags generated by the SVCE module, and how often our system missed intelligence because it discarded relevant details. We did not evaluate our entity matching process as it was done through DBpedia APIs. Human assessments and annotation was done by students familiar with the cybersecurity domain.

For our first evaluation measure we check the quality of tags generated by our SVCE module. We tagged 150 randomly selected security issues and then manually checked the tags. 
The annotators had to evaluate if the SVCE output was correct, partially correct or wrong. Our annotators agreed on the fact that 98 issues were marked correctly by the SVCE module and out of the remaining 52 issues, 18 were tagged completely wrong and the remaining were tagged partially correct. The annotators were then asked to look into the alerts generated for the 98 issues correctly identified, the system raised appropriate alerts for each. 


We evaluated the loss of intelligence because of discarded issues, i.e., those not included in the dataset of 9,194 security issues. A random sample of 200 issues was generated from the discarded issues. In these, our annotators found 9 issues with actionable security related information. We believe that these were wrongfully tagged by our SVCE module because of spelling mistakes, unidentifiable characters, informal slang expressions, non-English words, etc.

\section{Conclusion \& Future Work}\label{conc}

In this paper, we described our method for mining threat intelligence about open-source projects and libraries from host repository issues and bug reports. We collected bug reports consisting of all reported issues from multiple open source repositories on Github with Github's Rest API. Then, we extracted all the issues that had information related to cybersecurity. These security issues were then stored into our knowledge graph as RDF triples. We also track library and project dependencies for installed software on a client machine. The library dependencies with their corresponding applications are stored as RDF triples in the security knowledge graph. The knowledge graph was then used to create an alert and query system that can warn users of security vulnerabilities. Security analysts and developers can then query and receive alerts from the knowledge graph if any threat intelligence is found about linked libraries and projects, utilized in their products. The system can also issue alerts, about installed libraries, if new vulnerabilities are disclosed on these code repositories. 

For future work, we will like to mine threat intelligence from other web services, such as GitLab \cite{gitlab}, Code Triage \cite{triage}, etc, as they will also contain information about open-source projects with security vulnerabilities. These different code repositories host different projects which can serve as vital sources of threat intelligence. This variety would further improve our system and add more information to our knowledge graph. 
We would also like to investigate other knowledge representation techniques that can be used to store threat intelligence. Using multiple representations like vector space embeddings, can further improve the quality of applications built to utilize them.

\section*{Acknowledgement}
The work was partially supported by a gift from IBM Research, USA and the UMBC Louis Stokes Alliance for Minority Participation - LSAMP.

\bibliographystyle{plain}
\bibliography{bib,Bibliography,PHD}

\end{document}